# Enhanced coercive force of nanoparticles of special morphology in the Stoner-Wohlfarth model

Vladimir P. Savin,[1,*] and Yury A. Koksharov[1,2,3]

[1]Faculty of Physics, M.V.Lomonosov Moscow State University, Leninskie Gory, Moscow, 119991, Russia

[2]Kotelnikov Institute of Radio-Engineering and Electronics, Russian Academy of Sciences, Moscow, 125009 Russia

[3]Faculty of Physics and Mathematics, Shenzhen MSU-BIT, Shenzhen, 518179, China

*Contact author: svladimir9915@gmail.com

## ABSTRACT

We have found an unusual effect of increasing the coercive force in the Stoner-Wohlfarth model applied to magnetic nanoparticles of a special morphology. The particles consist of a ferromagnetic single-domain core surrounding by a magnetically soft shell. We have studied thoroughly individual and collective properties of these particles both through numerical calculations and analytical analysis. Fairly accurate approximate analytical formulas have been obtained for determining magnetization, hysteresis loop, coercive force, and other magnetic properties of the particles. The physical reason of the coercivity enhancement effect is the magnetic screening of a particle core by its shell. We have found the unambiguous conditions necessary for the existence of this effect. The large magnetization and moderate magnetic anisotropy of the core favor the effect of the coercivity enhancement.

# I. INTRODUCTION

Magnetic nanoparticles have been actively studied in recent decades due to their interesting fundamental properties and potentially important practical applications [1-12]. In a simplest model [13], a magnetic nanoparticle is a single ferromagnetic domain of spherical or ellipsoidal shape. This single-domain model makes it possible to explain the main features of the specific behavior of magnetic nanoparticle systems: superparamagnetism [14], the difference between ZFC-FC magnetic properties and the phenomenon of magnetic blocking [15]. A core-shell model has been proposed and experimentally realized [16-45]. In this model, magnetic properties of the inner part of the particle (core) and the outer part (shell) differ significantly that affects drastically on most magnetic characteristics, such as remnant magnetization and coercive force.

The coercivity and its mechanism in microscopic magnets and magnetic nanoparticles is a very important topic in magnetism. The coercivity of technical magnetics is much lower than the theoretical value [46]. The explanation of this paradox is rather complicate, since the coercivity mechanism depends on many microstructural parameters of materials. For example, it was found that the coercive force of sintered magnets depends on grain size and a reversion of the magnetization takes place by: (i) by thermally activated rotation processes in superparamagnetic particles smaller than 2 - 3 nm; (ii) a homogeneous rotation process in single-domain particles up to diameters of 10 nm; (iii) inhomogeneous rotation processes characterized by a decreasing nucleation field for particles up to several micrometers in diameter; (iv) domain wall displacements in multidomain grains. According to the work [46], for an ideal magnet each magnetically hard grain should be isolated from other grains by nonmagnetic grain boundary phase to suppress exchange coupling. On the other hand, possible increasing of the coercive force due to exchange coupling in so called "exchange-spring" composite structures, which contain both soft and hard magnetic phases, has been predicted and studied extensively [47-49]. Many experiments have been carried out and various variants of theoretical models have been introduced to realize and explain exchange-spring phenomena [35, 36, 50-63].

Nanoparticles with a core-shell morphology demonstrate unusual and controversial magnetic behavior [29, 31, 32, 37, 40, 41, 42, 57, 64]. Magnetic peculiarity can manifest itself in an increase of the coercive force for nanoparticles with a soft magnetic shell compared to nanoparticles without a shell [31, 32, 41, 42, 57, 62], a non-monotonic dependence of $H_c$ on the shell thickness [32, 37, 40, 41, 42], increasing of $H_c$ for nanoparticles with a magnetically soft core and a hard shell [25]. The exchange coupling between the core and shell is considered as a most possible explanation of these phenomena [65]. However, other physical mechanisms can be responsible for complex magnetic effects in core-shell nanoparticles. Although the maximum coercive force in core-shell nanoparticles has been usually observed for the hard antiferromagnetic shell, but it was found experimentally [31] and theoretically [63] that the soft shell also can increase $H_c$. The increase of the coercive field can be accompanied even by decreasing of exchange field and shell thickness [64].

What physical mechanisms, other than exchange coupling, can affect the magnetic properties of core-shell nanoparticles? An effect of magnetostatic dipolar interactions on magnetic behavior of spherical core-shell nanoparticles was investigated by a fast Monte Carlo method [66]. It was found that the coercive force is tunable by both dipolar and core-shell interface exchange interactions. The dipolar field can compete with the ferromagnetic core-shell exchange coupling. In case of the large magnetization of the shell its dipolar field may stabilize the core magnetic moment [66].

To study magnetostatic effects without the distorting influence of other factors we considered thoroughly a model of core-shell hard-soft (CSHS) nanoparticles without exchange coupling at a core-shell interface. In this model the core consists of single-domain hard magnet and the shell is simple soft magnet [67]. We used the Stoner-Wohlfarth model, which is an acceptable approach to evaluate the hysteretic behavior of various ferromagnetic systems [68], [69], including core-shell nanoparticles [70].

In the SW model the maximum value of the magnetic coercive force for a particular material can be achieved, corresponding to uniform remagnetization by the coherent rotation [71]. The coercive force in the SW model depends on the magnetic anisotropy field of the particle core and a direction of the external magnetic field [72, 73]. Our model of the CSHS nanoparticles [67] shows that the switching field, which reverses the core magnetic moment, depends on the magnetic screening effect produced by a particle shell. It can lead to an increase in a switching field and, as a result, to an enhancement of the coercive force.

## II. MODEL OF CSHS NANOPARTICLE

Let us consider a spherical CSHS nanoparticle in a nonmagnetic medium. The particle structure is illustrated in Fig. 1. The CSHS nanoparticle consists of a single-domain core with magnetic moment $\vec{m}_0$, magnetization $\vec{M}_0$ and uniaxial magnetic parameter anisotropy and a homogeneous soft magnetic shell with permeability $\mu$. The radius of the core is a, and the thickness of the shell is $d = b - a$, where b is the particle radius. The particle is placed in the external homogeneous field $\vec{H}_0$.

The magnetic moment $\vec{m}$ of the CSHS particle can be represented as:

$$\vec{m} = \vec{m}_0 + \vec{m}_{sh}, \qquad (1)$$

where $\vec{m}_{sh}$ is the magnetic moment of the shell.

The derivation of general formulas for the magnetic field in regions I-III (see Fig. 1) of the CSHS nanoparticle is presented elsewhere [67]. The magnetic moment of the shell $\vec{m}_{sh}$ and the total magnetic moment of the CSHS nanoparticle $\vec{m}$ can be written in the form (see Eqs. (28) and (33) in [67]):

$$\vec{m}_{sh} = (1 - A)\left(\frac{2\mu + 1}{2\mu - 2}b^3\vec{H}_0 - \vec{m}_0\right); \qquad (2)$$

$$\vec{m} = A\vec{m}_0 + (1 - A)\frac{2\mu + 1}{2\mu - 2}b^3\vec{H}_0, \qquad (3)$$

where

$$A = \frac{1}{1 + \frac{2}{9}\frac{(\mu - 1)^2}{\mu}\frac{(b^3 - a^3)}{b^3}} = \frac{1}{1 + \frac{2}{9}\frac{(\mu - 1)^2}{\mu}\frac{V_{sh}}{V_{CSHS}}} \qquad (4)$$

is the magnetic screening factor; $V_{sh} = 4\pi(b^3 - a^3)/3$ is the volume of the shell and $V_p = 4\pi b^3/3$ is the total volume of the CSHS particle. The factor $A$ depends only on the geometric ($b/a$) and magnetic ($\mu$) properties of the shell.

The magnetic anisotropy axis is directed along the $z$ axis (see Fig. 1). The core magnetic moment $\vec{m}_0$ is determined in a spherical coordinate system by equation:

$$\frac{\vec{m}_0}{m_0} = (\sin\theta \cos\varphi, \sin\theta \sin\varphi, \cos\theta). \qquad (5)$$

where $\theta$ and $\varphi$ are the polar and azimuth angles, respectively.

Similarly, the external magnetic field $\vec{H}_0$ is determined by $\theta_H$ and $\varphi_H$ (Fig. 1):

$$\frac{\vec{H}_0}{H_0} = (\sin\theta_H \cos\varphi_H, \sin\theta_H \sin\varphi_H, \cos\theta_H). \qquad (6)$$

A projection of $\vec{m}_0$ on $\vec{H}_0$ is equal to

$$m_{0\vec{H}_0} = \left(\vec{m}_0, \frac{\vec{H}_0}{H_0}\right) = m_0(\sin\theta \sin\theta_H \cos(\varphi - \varphi_H) + \cos\theta \cos\theta_H). \qquad (7)$$

The magnetic moment $\vec{m}_0$, the anisotropy z-axis and the external field $\vec{H}_0$ lie in the same plane as usual in the Stoner-Wohlfarth model. Therefore, $\varphi = \varphi_H$ and we can rewrite Eq. (7) as follows:

$$m_{0\vec{H}_0} = m_0 \cos(\theta - \theta_H). \tag{8}$$

Using (3) and (8), we get the expression for the magnetic moment $\vec{m}$ of the CSHS particle (see Eq. (3)) onto the direction of the external magnetic field $\vec{H}_0$:

$$m_{\vec{H}_0} = \left(\vec{m}, \frac{\vec{H}_0}{H_0}\right) = A m_0 \cos(\theta - \theta_H) + (1 - A)\frac{2\mu + 1}{2\mu - 2} b^3 H_0, \tag{9}$$

where the angle difference $\theta - \theta_H$ corresponds to the angle between $\vec{m}_0$ and the external field $\vec{H}_0$ (see Fig. 1).

## III. CSHS NANOPARTICLE IN THE SW MODEL

To consider the behavior of a magnetic particle in the SW model, we have to express the particles energy as the sum of three terms:

$$W = W_a + W_1 + W_2. \tag{10}$$

The first two terms correspond to the classical SW model [68, 72, 73, 74] and the third term appears due to the presence of a shell. The first term of Eq. (10) is the magnetic anisotropy energy:

$$W_a = K_a V_0 \sin^2 \theta, \tag{11}$$

where $K_a$ is the effective anisotropy constant, $V_0$ is the core volume, $\theta$ is the angle between $\vec{m}_0$ and the anisotropy axis (see Fig. 1 and Eq. (3)). The axis of magnetic anisotropy corresponds to the direction of easy magnetization ($K_a > 0$).

The second term of Eq. (10) is the interaction energy of the core magnetic moment $\vec{m}_0$ with the external homogeneous magnetic field $\vec{H}_0$:

$$W_1 = -(\vec{m}_0 \vec{H}_0) = -m_0 H_0 \cos(\theta - \theta_H). \tag{12}$$

The third term is the interaction energy of the shell with the external field $\vec{H}_0$. We have obtained the following expression (see [67]):

$$W_2 = -\frac{1}{2} \int_{V_{Sh}} \left(\vec{M}_{sh}(\vec{H}_0 + \vec{H}_{core})\right) dV =$$

$$= -\frac{1}{2}(1-A)\left[\frac{2\mu+1}{2\mu-2}b^3 H_0^2 - 2(\vec{m}_0\vec{H}_0) + \frac{\mu+2}{\mu-1}\frac{m_0^2}{a^3}\right], \tag{13}$$

where $\vec{H}_{core}$ is the magnetic field of the core, $\vec{M}_{sh}$ is the magnetization of the shell. The integration is extended over the volume of the shell.

Introducing the dimensionless magnetic field $h_0$ and the magnetic anisotropy field $H_a$:

$$h_0 = \frac{H_0}{H_a}; \tag{14}$$

$$H_a = \frac{2K_a V_0}{m_0} = \frac{2K_a}{M_0}; \tag{15}$$

as well as the parameter $B$, which is not depend on any angles:

$$B = (1-A)\frac{1}{2}\left[\frac{2\mu+1}{2\mu-2}\frac{b^3 H_a}{m_0}h_0^2 + \frac{\mu+2}{\mu-1}\frac{m_0}{a^3 H_a}\right], \tag{16}$$

we can rewrite the energy (10) in dimensionless form:

$$w = \frac{W}{2K_a V_0} = \frac{1}{2}\sin^2\theta - Ah_0\cos(\theta-\theta_H) - B. \tag{17}$$

If there is no shell ($A = 1$, $B = 0$), Eq. (17) takes the well-known form [73]:

$$w = \frac{1}{2}\sin^2\theta - h_0\cos(\theta-\theta_H). \tag{18}$$

Figure 2 shows the dependencies of $A$ versus $\mu - 1$ for different ratios $b/a$. The inset in Fig. 2 presents the curves $B$ versus $A$ for fixed $b/a$ and different values of $H_a/M_0$, $H_0/H_a$. The parameter $A$ decreases from 1 ($\mu = 1$) to zero ($\mu \gg 1$) for any value of $b/a > 1$. In Eq. (16), the expression in square brackets is implicitly depends on $A$. The value of $B$ decreases linearly with increasing $A$, if $A \lesssim 0.8$, and tends to zero, if $A \to 1$ (see the inset in Fig. 2).

The task of the SW model is to find the value $\theta = \theta^*$, that corresponds to the local minimum of the energy in Eq. (17). At low magnetic fields we can distinguish two local minima corresponding to stable and metastable states (Fig. 3). The stable state corresponds to the absolute minimum of the energy. If $h_0 \geq h_{sw}$, the energy minimum is unique, where $h_{sw}$ is the switching field, which depends on the value of $\theta_H$. This dependence is described by the SW astroid for shell-less single-domain particles [73].

The dependence of the energy (17) on the angle $\theta$ is shown in Fig. 3(a, b) for the longitudinal ($\theta_H = 0$) and transverse ($\theta_H = \pi/2$) field, respectively. The cases of $A = 0.5$ and $A = 1$ (no shell) are presented. The conditions for energy extremes allow us to obtain universal expressions for the values of energy barriers $\Delta w_1$ and $\Delta w_2$ (see Fig. 3(a, b)):

$$\Delta w_1 = \frac{1}{2}(1 + Ah_0)^2; \tag{19}$$

$$\Delta w_2 = \frac{1}{2}(1 - Ah_0)^2. \tag{20}$$

Equations (19), (20) show that the energy barriers between local minima disappears, if $h_0 = 1/A$. This is true for both cases $\theta_H = 0$ and $\theta_H = \pi/2$.

To find the expression of the astroid for arbitrary directions of $\vec{H}_0$, we used the algorithm of Bertotti [73] for a particle without a shell. The field components perpendicular and parallel to a simple axis can be written as:

$$h_\perp = h_0 \sin \theta_H; \tag{21}$$

$$h_\parallel = h_0 \cos \theta_H. \tag{22}$$

Using Eqs. (21)-(22), it is convenient to rewrite the Eq. (17) in the form:

$$w = \frac{1}{2} \sin^2 \theta - Ah_\parallel \cos \theta - Ah_\perp \sin \theta - B. \tag{23}$$

The switching field is achieved, if both the first and second derivatives with respect to $\theta$ are equal to zero. Equating to zero derivative $\partial w/\partial \theta$, we obtained the equation:

$$\frac{\partial w}{\partial \theta} = \sin \theta \cos \theta + Ah_\parallel \sin \theta - Ah_\perp \cos \theta = 0. \tag{24}$$

Then, equating to zero the second derivative $\partial^2 w/\partial \theta^2$, we found the following equation:

$$\frac{\partial^2 w}{\partial \theta^2} = \cos^2 \theta - \sin^2 \theta + Ah_\parallel \cos \theta + Ah_\perp \sin \theta = 0. \tag{25}$$

Solving the set of Eqs. (24)-(25) with respect to $h_\parallel$ and $h_\perp$, we obtained:

$$Ah_\parallel = -\cos^3 \theta; \tag{26}$$

$$Ah_\perp = \sin^3 \theta. \tag{27}$$

Raising Eq.(26), (27) to the power of 2/3 and then adding them up, we got:

$$(Ah_\perp)^{2/3} + (Ah_\parallel)^{2/3} = 1. \tag{28}$$

Substituting Eqs. (21), (22) into Eq. (28), it is easy to obtain the desired expression for the astroid, which determines the switching field for any value of $\theta_H$:

$$h_{sw} = \frac{1}{A(\sin^{2/3}\theta_H + \cos^{2/3}\theta_H)^{3/2}}. \tag{29}$$

It follows from Eq. (29) that the switching field for the CSHS nanoparticles increases by a factor of $1/A$ compared to single-domain nanoparticles without the shell. This indicates that the external magnetic field is screened by the CSHS particle shell.

### IV. HYSTERESIS LOOPS IN THE SW MODEL

Let us now consider of an ensemble of identical noninteracting CSHS nanoparticles with arbitrarily oriented anisotropy axes in the external homogeneous field $\vec{H}_0$. In order to obtain hysteresis loops, it is necessary to average the projection of the magnetic moment of the particle onto the direction of the external magnetic field (over the angle $\theta_H$), assuming that the anisotropy axes distributed at random:

$$\frac{\langle m_{\vec{H}_0}\rangle}{m_0} = \frac{\int_0^{2\pi} d\varphi_H \int_0^{\pi/2} \frac{m_{\vec{H}_0}}{m_0} \sin\theta_H \, d\theta_H}{\int_0^{2\pi} d\varphi_H \int_0^{\pi/2} \sin\theta_H \, d\theta_H} = \int_0^{\pi/2} \frac{m_{\vec{H}_0}}{m_0} \sin\theta_H \, d\theta_H. \tag{30}$$

Taking into account (9) we can rewrite the Eq. (30) as follows:

$$\frac{\langle m_{\vec{H}_0}\rangle}{m_0} = A \int_0^{\pi/2} \cos(\theta^* - \theta_H) \sin\theta_H \, d\theta_H + (1-A)\frac{2\mu+1}{2\mu-2}b^3 \frac{H_0}{m_0}, \tag{31}$$

where $\theta^*$ is determined from Eq. (17) and the energy minimum condition:

$$\left(\frac{\partial w}{\partial \theta}\right)\bigg|_{\theta=\theta^*} = 0; \quad \left(\frac{\partial^2 w}{\partial \theta^2}\right)\bigg|_{\theta=\theta^*} > 0. \tag{32}$$

It is important to note that: (i) $\theta^*$ depends on $\theta_H, H_0, M_0, K_a$; (ii) Eq. (31) does not depend on the particle size, it depends only on the $b^3/a^3$ ratio; (iii) the second term in Eq. (31) increases linearly in the field $H_0$ and does not depend on the magnetic anisotropy $H_a$.

Depending on the magnitude of $h_0$, there may be up to two local minima $\theta^*$ in the interval $[0, \pi]$. The selection of the desired $\theta^*$ is determined by the magnetic history. Let us assume that $h_0 > 0$ ($\theta_H \in [0; \pi/2]$) acts on the particles, which have been magnetized by $h_0 < 0$ and $\theta \in [\pi/2; \pi]$. If $h_0 < h_{sw}$ then $\theta^* \in (\pi/2; \pi]$. This angle is related to not absolute minimum of the energy for $h_0 > 0$. If $h_0 > h_{sw}$ then $\theta^* \in [0, \pi/2]$. This angle is related to absolute energy minimum for $h_0 > 0$.

Figure 4 shows the hysteresis loops with different parameters $\mu$ and fixed $b/a = 1.1$. In the intervals $[-0.5/A; 0.75/A]$ and $[-0.75/A; 0.5/A]$, the dotted curves correspond to an analytical formular:

$$\frac{\langle m_{\vec{H}_0} \rangle}{m_0} = A\left[\frac{1}{3}(Ah_0)^3 \mp \frac{1}{2}(Ah_0)^2 + \left(\frac{2}{3} + \frac{3}{4\pi}\frac{(1-A)}{A^2}\frac{2\mu+1}{2\mu-2}\frac{b^3}{a^3}\frac{H_a}{M_0}\right)Ah_0 \pm \frac{1}{2}\right]. \quad (33)$$

If $h_0 \in [0.75/A; \infty]$ and $h_0 \in [-\infty, -0.75/A]$, then the upper and lower branches can be expressed by the following approximate expression:

$$\frac{\langle m_{\vec{H}_0} \rangle}{m_0} = \pm A\left(1 - \frac{1}{15}\frac{1}{(Ah_0)^2} - \frac{1}{315}\frac{1}{(Ah_0)^4} \pm \frac{3}{4\pi}\frac{(1-A)}{A^2}\frac{2\mu+1}{2\mu-2}\frac{b^3}{a^3}\frac{H_a}{M_0}Ah_0\right). \quad (34)$$

If $h_0 \in [-0.75/A; -0.5/A]$ and $h_0 \in [0.5/A, 0.75/A]$, then the following approximate expression is valid:

$$\frac{\langle m_{\vec{H}_0} \rangle}{m_0} = \pm A\left(\frac{1}{(Ah_0)^2} - 1\right)\ln\left[(1 \mp Ah_0)\left(\frac{2-(Ah_0)^2 + \sqrt{(Ah_0)^4 - 4\left(\frac{1-(Ah_0)^2}{3}\right)^3}}{2-(Ah_0)^2 - \sqrt{(Ah_0)^4 - 4\left(\frac{1-(Ah_0)^2}{3}\right)^3}}\right)\right] +$$

$$+ \frac{1}{h_0} + \frac{3}{4\pi}\frac{(1-A)}{A}\frac{2\mu+1}{2\mu-2}\frac{b^3}{a^3}\frac{H_a}{M_0}Ah_0. \quad (35)$$

In Eqs. (33), (34), (35) the upper and lower signs are related to upper and lower branches, respectively.

Equations (33-35) were obtained by a method similar to the one used in the work [75]. Surprisingly, these approximate analytic formulas demonstrate a good agreement with numerical calculations by using Eq. (31).

As $\mu$ increases ($A$ decreases, see Eq. (4) and Fig. 2), the hysteresis loop lengthens and the remanent magnetic moment $m_r$ decreases. By finding $\theta^*$ from Eq. (32) and substituting the resulting value of $\theta^*$ and $H_0 = 0$ in Eq. (31), we can obtain a simple analytical expression for $m_r$:

$$m_r = \langle m_{\vec{H}_0} \rangle\big|_{H_0=0} = \pm\frac{A}{2}m_0. \quad (36)$$

The same result can be obtained from Eq. (33) by taking $h_0 = 0$.

The conjugate point of Eqs. (33), (35) corresponds to a kink of the hysteresis branches (see Fig. 4). As $\mu$ increases (A decreases), kink points become more noticeable on the hysteresis loops. The kink points arise when $h_0$ reaches the minimal value of $h_{sw}$ relative to $\theta_H$, which is the same for all $A$ and corresponds to the angle $\theta_H = \pi/4$ (see Eq. (29)). From Eq. (29), we found a formula for the field $H_k$ (or $h_k$ in normalized form) that relates to the kink point on a hysteresis loop:

$$h_k = \frac{H_k}{H_a} = \pm\frac{1}{2A}. \tag{37}$$

In the next section the analytical formula for the coercive force will be obtained and discussed.

## V. ENHANCED COERCIVE FORCE IN THE SW MODEL

The formula (33) for the average magnetic moment can be used to find analytical expression for the coercive force of the system by taking $\langle m_{\overline{H}_0}\rangle = 0$:

$$\frac{1}{3}(Ah_c)^3 + \frac{1}{2}(Ah_c)^2 + \left(\frac{2}{3} + \frac{3}{4\pi}\frac{(1-A)}{A^2}\frac{2\mu+1}{2\mu-2}\frac{b^3}{a^3}\frac{H_a}{M_0}\right)Ah_c - \frac{1}{2} = 0 \tag{38}$$

Figure 5 shows the dependence of the dimensionless coercive force $h_c = H_c/H_a$ on the parameter $A$ for various core magnetic materials. The solid and dashed curves were drawn numerically by Eq. (31) and analytically by Eq. (38), respectively. The dependences in Fig. 5 demonstrate that there can be a maximum of the coercive force $h_{c,max}$ at some values of $A = A_{max}$. Both values $h_{c,max}$, $A_{max}$ depend on the core material parameter $H_a/M_0$ which affects the third term in Eq. (38).

Since $Ah_c$ is a variable in Eq. (38) we can expect that $h_c$ increases as $A$ decreases and became larger then $h_c = 0.48$ for shell-less particles. We called this effect the enhancement of the coercive force for CSHS nanoparticles. In Fig. 5, this effect (the condition $h_c > 0.48$) takes place for all materials shown with the exception of cobalt. The effect of the enhancement of $h_c$ is most pronounced for $\alpha$-Fe due to the favorable combination of material parameters, as discussed below.

The existence of a limit value $H_{c,sup}$ for a given $H_a/M_0$ is demonstrated in Figs. 6, 7. Figures 6(a, b) show the dependences of $H_c$ on $b/a$ for iron and cobalt cores with different values of $\mu$. We can see that $H_{c,max} = h_{c,max}H_a$ increases if $b/a \to 1$ and tends to constant value $H_{c,sup}$ at large values of $\mu$ and some value of $A = A_{sup}$. The inserts in Figs. 6(a, b) demonstrate the convergence of values $A_{max}$ and $H_{c,max}$ to their supremum values $A_{sup}$ and $H_{c,sup}$, respectively, with increasing $\mu$. The value $H_{c,sup}$ corresponds to the maximum of $H_c$ that can be achieved by using a magnetically soft shell for a given core material. For example, in the cases of Fe and Co core particles the supremum values of coercive force $H_{c,sup}$ are 660 Oe and 3480 Oe, respectively (Fig. 6(a, b)).

The dependencies $H_c$ versus $\mu$ for different values of $b/a$ are shown for Fe and Co core particles (Fig. 7(a, b)). The local maximum on $H_c(\mu)$ curve disappears with an increase $b/a$. This is consistent with the dependence $h_c(A)$ for the Co core particles with $b/a = 1.1$ (Fig. 5). The same maximum values of $H_{c,sup}$ detected in Fig. 6(a, b) are also observed in Fig. 7(a, b).

We can obtain the values $h_{c,sup}$ and $A_{sup}$ analytically by using Eq. (38). It is convenient to rewrite Eq. (38) as follows:

$$x^3 + \frac{3}{2}x^2 + \beta x - \frac{3}{2} = 0, \tag{39}$$

where $x = Ah_c$ and

$$\beta = 2 + \frac{9}{4\pi}\frac{(1-A)}{A^2}\frac{2\mu+1}{2\mu-2}\frac{b^3}{a^3}\frac{H_a}{M_0}. \tag{40}$$

The solution of Eq. (39) is given by the Cardano formula:

$$x = \sqrt[3]{-q/2 + \sqrt[2]{\Delta}} + \sqrt[3]{-q/2 - \sqrt[2]{\Delta}} - \frac{1}{2}, \tag{41}$$

where

$$q = \frac{1}{4} - \frac{\beta}{2};\ p = \beta - \frac{3}{4};\ \Delta = \left(\frac{p}{3}\right)^3 + \left(\frac{q}{2}\right)^2.$$

The maximum value of $x$ relates to the minimal value of $\beta$ in Eq. (39). It can be proved as follows:

$$\frac{dx}{d\beta} = -\frac{x}{3x^2 + 3x + \beta} < 0,$$

where $x > 0$, $\beta > 0$.

The minimal value of $\beta$ is equal to 2, if $A = 1$ (see Eq. (40)). In case of $A < 1$ the minimal value of $\beta$ can be achieved by taking $b \to a$ and $\mu \gg 1$. Using these conditions we can rewrite Eq. (38) as

$$(Ah_c)^3 + \frac{3}{2}(Ah_c)^2 + \left(2 + \frac{9}{4\pi}\frac{(1-A)}{A^2}\frac{H_a}{M_0}\right)Ah_c - \frac{3}{2} = 0. \tag{42}$$

The properties of the shell are revealed in Eq. (42) only by the screening factor $A$. The solution of Eq. (42) can be obtained by using Eq. (41) taking into account the modified expression for $\beta$ (see Eq. (40)):

$$\beta = 2 + \frac{9}{4\pi}\frac{(1-A)}{A^2}\frac{H_a}{M_0}. \tag{43}$$

Since parameter $A_{max}$ saturates (see inserts in Fig. 6(a, b)) it is convenient to use parameter $A$ instead of $\mu$ or $b/a$ to display the limiting coercivity curves force with a local maximum at $H_{c,sup}$. Equation Eq. (42) determines the limiting curves $h_c(A)$ for fixed ratio $H_a/M_0$, which are shown in unnormalized form $H_c(A)$ in Fig. 8(a) and Fig. 9(a) for different $K_a$ (fixed $M_0$) and for different $M_0$ (fixed $K_a$), respectively. Solid lines in Figs. (8a, 9a) are drown by using the Cardano's solution of the cubic equation (42). Dashed lines in Figs. (8a, 9a) present the analytical dependencies $H_{c,sup}$ versus $A_{sup}$ that were obtained as follows.

Firstly, the derivative of Eq. (42) in relation to A should be equal zero in case of $h_c = h_{c,sup}$ and $A = A_{sup}$. So, we have found an equation that links $h_{c,sup}$ и $A_{sup}$ for fixed ratio $H_a/M_0$:

$$A_{sup}^2 h_{c,sup}^2 + A_{sup} h_{c,sup} - \frac{3}{4\pi A_{sup}^2}\frac{H_a}{M_0} + \frac{2}{3} = 0. \tag{44}$$

The general equation (42) is also valid for $A = A_{sup}$ and $h_c = h_{c,sup}$:

$$\left(A_{sup} h_{c,sup}\right)^3 + \frac{3}{2}\left(A_{sup} h_{c,sup}\right)^2 + \left(2 + \frac{9}{4\pi}\frac{(1-A_{sup})}{A_{sup}^2}\frac{H_a}{M_0}\right) A_{sup} h_{c,sup} - \frac{3}{2} = 0. \tag{45}$$

The set of Eqs. (44)–(45) have three independent variables $A_{sup}$, $h_{c,sup}$, $H_a/M_0$ and can be used to find the relations between them. It is convenient to obtain the relation between $H_a/M_0$ and $A_{sup}$. To do this, we need to find the unique positive root of square equation (44):

$$h_{c,sup} = \frac{1}{2A_{sup}}\left(\sqrt{\frac{3}{\pi A_{sup}^2}\frac{H_a}{M_0} - \frac{5}{3}} - 1\right). \tag{46}$$

Substituting Eq. (46) to Eq. (45) we obtain the cubic algebraic equation with respect to $H_a/M_0$:

$$\frac{27}{\pi^3}\frac{(4 - 3A_{sup})^2}{A_{sup}^6}\left(\frac{H_a}{M_0}\right)^3 - \frac{9}{\pi^2}\frac{(24A_{sup}^2 - 38A_{sup} + 9)}{A_{sup}^4}\left(\frac{H_a}{M_0}\right)^2 +$$

$$+\frac{424}{\pi}\frac{(A_{sup}-1)}{A_{sup}^2}\frac{H_a}{M_0}-\frac{9248}{27}=0. \tag{47}$$

Equation (47) can be solved analytically by using the Cardano's formula, which gives expression for $H_a/M_0$ as a function of $A_{sup}$:

$$\frac{H_a}{M_0}=\frac{4\pi}{3}\left[\sqrt[3]{-q_1/2+\sqrt[2]{\Delta_1}}+\sqrt[3]{-q_1/2-\sqrt[2]{\Delta_1}}+\frac{(9-38A_{sup}+24A_{sup}^2)}{12(4-3A_{sup})^2}A_{sup}^2\right]; \tag{48}$$

where

$$p_1=-\frac{A_{sup}^4(6865-17644A_{sup}+15868A_{sup}^2-5640A_{sup}^3+576A_{sup}^4)}{48(4-3A_{sup})^4};$$

$$q_1=-\frac{A_{sup}^4}{864(4-3A_{sup})^6}\left[1276057-4176114A_{sup}+5439792A_{sup}^2-3561308A_{sup}^3+\right.$$

$$+1215288A_{sup}^4-203040A_{sup}^5+13824A_{sup}^6];$$

$$\Delta_1=\left(\frac{p_1}{3}\right)^3+\left(\frac{q_1}{2}\right)^2;$$

Substituting Eq. (48) back into Eq. (46) we obtain the cumbersome formula for the function $h_{c,sup}(A_{sup})$. This formula is universal because it does not depend on the properties of the core. This universal dependence is shown in Fig.8.

Dashed lines in Figs. 9(a)-10(a) intersect the limiting curves $H_c(A)$ at maximum points $H_{c,sup}(A_{sup})$. The values $H_{c,sup}$ have been calculated using Eqs. (46), (48) and representing $K_a$ and $M_0$ as a function of $A_{sup}$ Eq. (48).

From Eq. (44), taking $A_{sup}=1$ and $A_{sup}h_{c,sup}=0.5$, the condition for the existence of a local maximum can be written as:

$$K_a\leq\frac{17\pi}{18}M_0^2 \tag{49}$$

It follows from Eqs. (46), (49) that if $M_0$ is fixed and $A_{sup}\to 1$, then $H_{c,sup}\to 17\pi M_0/18$ (see Fig. 9(a)). If $K_a$ is fixed and $A_{sup}\to 1$, then $H_{c,sup}\to\sqrt{17\pi K_a/18}$ (see Fig. 10(a)).

Figures 9a, 10a present the limiting coercivity curves $H_c(A)$ at different values of $K_a$ and fixed $M_0$ (Fig. 9a) or at different values of $M_0$ and fixed $K_a$ (Fig. 10a). An increase in $K_a$ (Fig. 9a)

or a decrease in $M_0$ (Fig. 10a) leads to the disappearance of the local maximum of $H_c(A)$. It means that the coercivity enhancement effect disappears if Eq. (49) is not valid.

Figures 9(b), 10(b) illustrate in detail the effect of the coercivity force enhancement that can be quantify by the difference $\Delta H_{c,sup} = H_{c,sup} - H_{c,sw}$. Dotted and solid lines represent numerical (Eq. (31)) and analytical calculations (Eq. (42)), respectively. Fig. 9b shows the dependencies of $H_{c,sup}$ and $A_{sup}$ versus $K_a$ if $M_0$ is fixed. The dashed line presents $H_{c,sw}(K_a)$ for a SW particle without the shell ($A = 1$). The dependencies $A_{sup}(K_a)$ and $H_{c,sup}(K_a)$ are monotonous, but the function $\Delta H_{c,sup}(K_a)$ demonstrate a local maximum (the inset in Fig. (9b)).

Fig. 10b shows $H_{c,sup}(M_0)$ and $A_{sup}(M_0)$ with fixed $K_a$. The dashed line presents the dependence $H_{c,sw}(M_0)$ for a SW particle. There is a good correspondence between numerical and analytical calculations by Eq. (31) and Eq. (42), respectively. According to inset in Fig. 9b the coercivity enhancement effect is stronger at higher values of $M_0$. An increase in $M_0$ leads to a decrease in both $H_{c,sup}$ and $A_{sup}$.

Fig. 11(a) shows the three-dimensional dependence of $\Delta H_{c,sup}$ on $M_0$ and $K_a$. Fig. 11(b) presents the dependence of $\Delta(H_{c,sup} M_r)$ on $M_0$ and $K_a$, where the remnant magnetization $M_r$ is equal to (see Eq. (36)):

$$M_r = \frac{A_{sup}}{2} \left(\frac{a}{b}\right)^3 M_0 \qquad (50)$$

and

$$\Delta(H_{c,sup} M_r) = H_{c,sw} M_0/2 - H_{c,sup} M_r \qquad (51)$$

The value of $H_c M_r$ is correlated with to the magnetic energy product [72]. The data in Fig. 11b prove that the value of $H_{c,sup} M_r$ for the CSHS particle cannot exceed the same parameter for the SW particle with the same core. The dotted parabolic line $K_a = 17\pi M_0^2/18$ (see Eq. (49)) delimits aria in which the enhancement effect occurs ($\Delta H_{c,sup} > 0$). Arrows in Fig. 11(a) indicate the possibility for realization of this effect for different materials.

## V. DISCUSSIONS

Our results shows that the main properties predicted by the SW model for a single-domain particle without a shell are also manifested for systems of nanoparticles with a magnetically soft shell. Interestingly, other approaches, such as micromagnetic simulations [70], also show the applicability of SW model formulas to core-shell magnetism problems. Certainly, the angular dependences for energy, the SW astroid equation, and hysteresis loops, both averaged and individual (for fixed mutual directions of the anisotropy axes and $H_0$) are modified for CSHS nanoparticles, but approximate formulas derived for CSHS nanoparticles are also valid for the standard Stoner-Wohlfarth model (see hysteresis loops in Fig. 4).

The shape of the astroid remains unchanged, but its size increases isotropically by a factor of $1/A$ (see Eq. 29). So, the switching field increases for all external field directions. The individual and averaged hysteresis loops are characterized by a decrease in the residual magnetization by a factor of $1/A$ (see Eqs. 9, 36). Thus, the parameter $A$ can be called a magnetic screening factor for CSHS particles (see also [67]). The counterintuitive relative increase in $H_c$ for particles with a shell compared to the same particles without a shell can also be explained by the screening effect. Indeed, the magnitude of a value of the switching field is determined by the astroid. The increase of the astroid size leads to respective increasing of the magnitude of the external magnetic field, which reverses the particle magnetization. The effect of an increase of the coercive force was proved experimentally [31, 32, 41, 42, 57, 62].

It is important to note that, firstly, not all values $K_a$ and $M_0$ provide the coercivity enhancement effect (see Fig. 11(a) and Eq. (49)). Low values of $K_a$ and high values of $M_0$ favor this effect, which is maximal for a small shell thickness. Decreasing of $H_c$ is observed for large values $\mu$ and $b/a$ (see Figs. 6, 7) for any magnetic material of the core. Secondly, as the coercive force increases, the residual magnetization decreases so that the product $H_c M_r$ remains less than the corresponding value for particles without a shell (Fig. 11(b)). Thirdly, our model assumes the absence of an exchange coupling between the core and the shell. It is possible to eliminate the exchange interaction between the core and the shell by using a non-magnetic layer between them. Experiments and theoretical considerations show that in hard/soft multilayers the introduction of a non-magnetic layer between the hard and soft phases can prevent a decrease in the coercive force [48].

In our model, the magnetic permeability is constant, which is an idealization of a real soft bulk magnet where magnetic permeability depends on the external magnetic field. Fortunately, soft nano-materials show a linear field dependence of the magnetization for rather large interval of fields compared with similar bulk materials [78-83].

The model of an absolutely hard magnetic core is also an idealization. Due to the core shell interactions (exchange coupling [63] or dipolar [66]) the shell can induce the inhomogeneous magnetization in the core. This effect also can lead to increasing of $H_c$ in core-shell nanoparticles in comparison with particles without a shell (see inset at Fig. 4 in [63]).

In any case, the SW model can qualitatively explain the possible physical causes of changes in the coercive force due to the interaction of the core and shell. The screening effect can both increase and decrease the coercive force, while the exchange coupling preferably reduces it. Some diversity in the experiments can be, probably, explained by the competition of these two mechanisms.

## VI. CONCLUSIONS

We analyzed magnetic properties of CSHS nanoparticles in the Stoner-Wohlfarth model. Numerical calculations have been supported by approximate analytical formulas for all individual and collective magnetic properties of the particles. For some core materials, averaged hysteretic loops show an increase of coercive force, which coincides with a decrease in the remnant magnetization. The effect of increasing the coercive force means that nanoparticles with a magnetically soft shell have a higher $H_c$ value compared to nanoparticles without a shell. There is a border in the plane $(K_a, M_o)$, indicating the values of the core magnetization $M_o$ and magnetic anisotropy constant $K_a$, which are in principle acceptable for the existence of the coercivity enhancement effect (see Eq. (49) and Fig. 11(a)). Unfortunately, most advanced hard magnetics, like Nd-Fe-B and Sm-Co, are not able to demonstrate this effect. Equation (49) and Fig. 11(a) specify that low values of $K_a$ and high values of $M_0$ are preferred for the coercivity enhancement effect. This mostly applies to iron oxides, which are popular materials in nanotechnology.

Experimental works indicate the possibility of the effect of an increase in the coercive force for particles with a magnetically soft shell. A direct comparison of our model with the experimental data from this work is not entirely correct, since the magnetic parameters of the core and shell for nanoparticles may differ from the corresponding volume values. Besides, the SW model does not take into account temperature fluctuations and interparticle interaction. In the future, we plan to consider these factors in an advanced model of a core-shell magnetic nanoparticle.

## ACNOWLEDGEMENTS

## DATA AVAILABILITY

No data were created or analyzed in this article.

## CONFLICT OF INTEREST

The authors declare no competing financial interest.

___

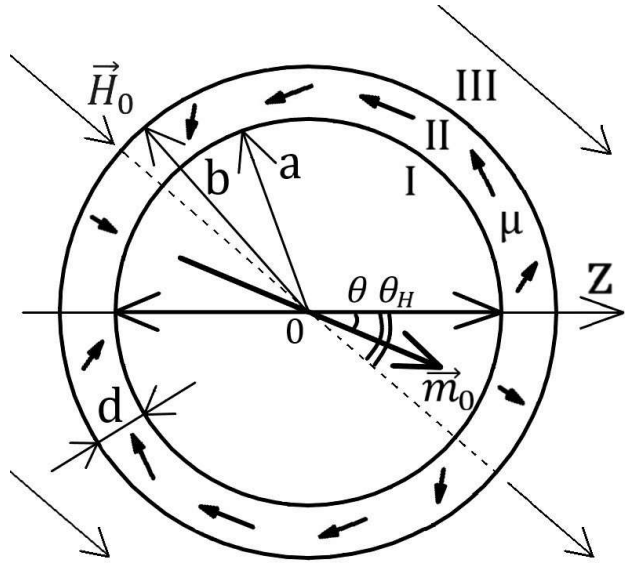

FIG. 1. Model of spherical CSHS nanoparticle with uniaxial core anisotropy in the external homogeneous magnetic field $\vec{H}_0$. The particle consists of a single-domain hard ferromagnetic core (region I) of radius $a$ with uniaxial anisotropy along z-axis, and a homogenous soft magnetic shell (region II) of external radius $b$ (thickness $d$) and magnetic permeability $\mu$. The particle is located in a nonmagnetic medium (region III). $\theta$ is the angle between magnetic moment of the core $\vec{m}_0$ and the z-axis, and $\theta_H$ is the angle between the external field vector $\vec{H}_0$ and z-axis. Small bold arrows inside the shell indicate its non-homogeneous magnetization (see Fig. 2 in [67]).

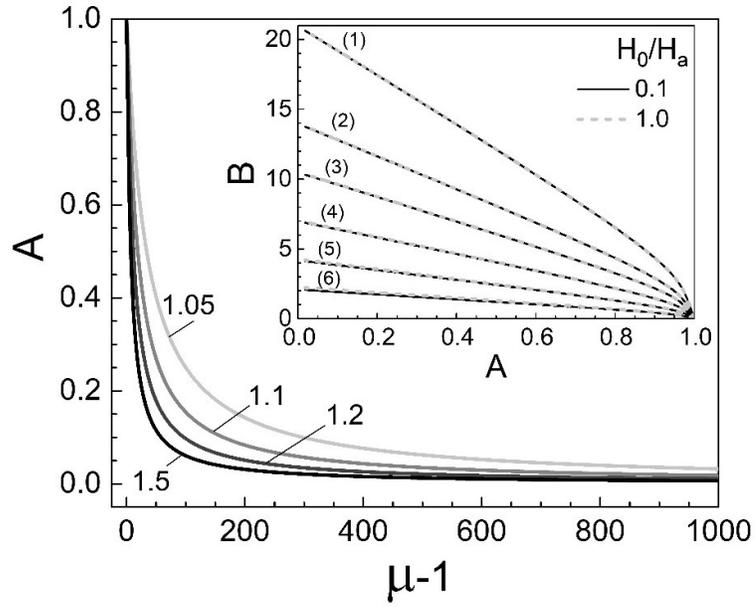

FIG. 2. The parameter $A$ (Eq. (4)) as a function of magnetic permeability $\mu$ for different values of ratio $b/a$. The values near to the curves indicate the b/a ratio. The inset shows the dependence of parameter $B$ (Eq. (16)) on $A$ with a fixed ratio $b/a = 1.1$ and different ratios $H_a/M_0$, $H_0/H_a$. The solid and dotted lines correspond to $H_0/H_a = 0{,}1$ and $H_0/H_a = 1{,}0$, respectively. The lines (1) – (6) were calculated by using $H_a/M_0$: (1) 0.1; (2) 0.15; (3) 0.2; (4) 0.3; (5) 0.5; (6) 1.0.

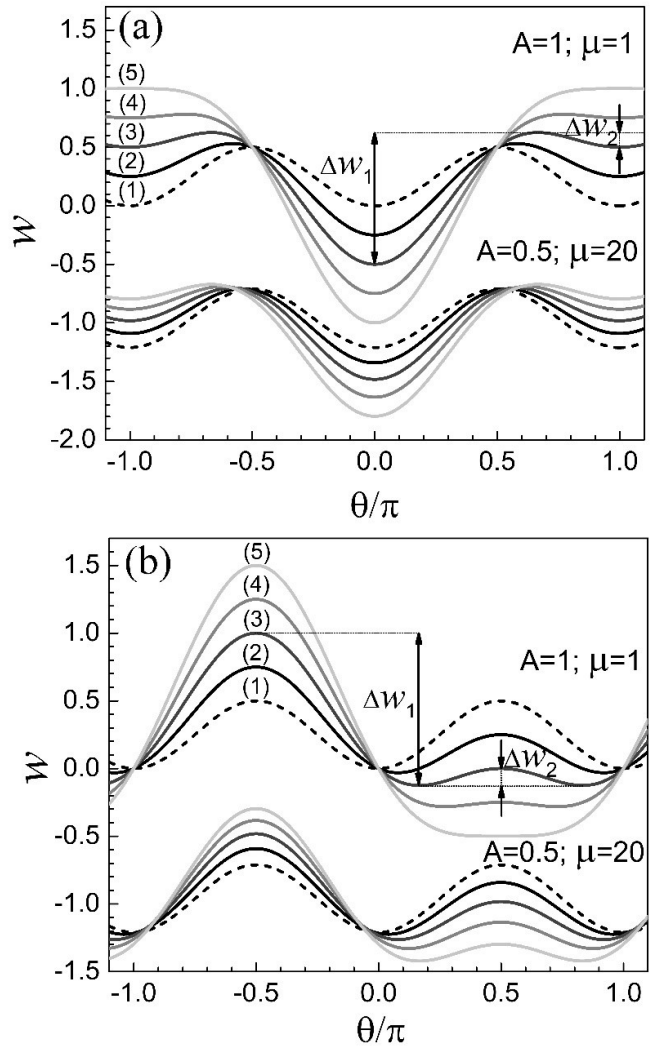

FIG. 3. Variations of the energy $w$ as a function of angle $\theta$ for different values of the dimensionless magnetic field $h_0$ (Eq. (17)). The external field is directed: (a) along the anisotropy axis ($\theta_H = 0$), (b) perpendicular to the anisotropy axis ($\theta_H = \pi/2$). The upper curves correspond to the case of the particle without the soft magnetic shell ($A = 1$; $\mu = 1$). The lower curves correspond to the CSHS particle for the cases $A = 0.5$; $\mu = 20$. The lines (1) – (5) were calculated by using $b = 1.1a$; $a = 10\ nm$; $H_A = M_0 = 1.7\ kG$; $h_0$: (1) 0; (2) 0.25; (3) 0.5; (4) 0.75; (5) 1.0.

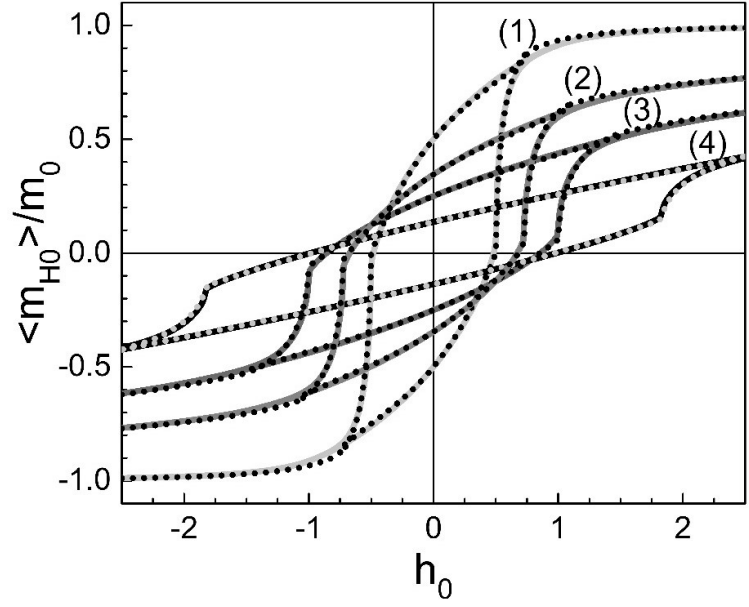

FIG. 4. The dimensionless average magnetic moment along the direction of the external field $\langle m_{\vec{H}_0}\rangle/m_0$ as a function of dimensionless external field strength $h_0$ (hysteresis loop) for different values of $\mu$ (parameter $A$). The lines (1) – (4) were calculated by using $b = 1.1a$; $a = 10\ nm$; $M_0 = 1.7\ kG$; $K_a = 0.5\ Merg/cm^3$ [74]; (1): $\mu = 1$ ($A = 1$); (2): $\mu = 10$ ($A = 0.7$); (3): $\mu = 20$ ($A = 0.5$); (4): $\mu = 50$ ($A = 0.27$). The line (1) corresponds to a nanoparticle without the soft magnetic shell (SW loop). The solid and dotted lines represent numerical (Eq. (31)) and analytical (Eqs. (33)-(35)) calculations, respectively.

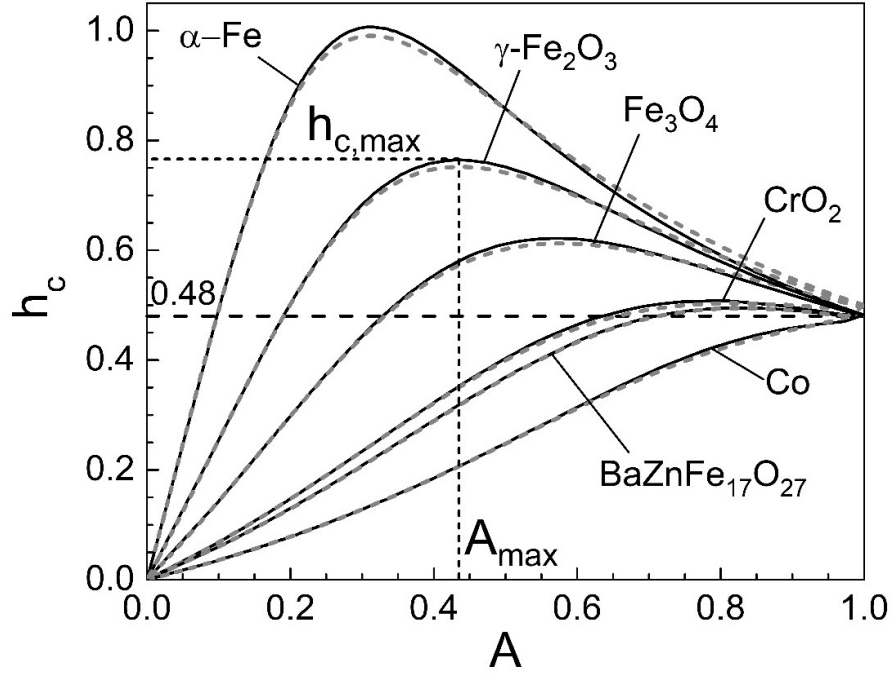

FIG. 5. The dimensionless coercive force $h_c$ as a function of $A$ at a fixed ratio $b/a = 1.1$; $a = 10\ nm$; for different core materials: α-Fe: $M_0 = 1710\ G$; $K_a = 0.48\frac{Merg}{cm^3}$ [74]; Co: $M_0 = 1440\ G$; $K_a = 5\ Merg/cm^3$ [74]; $CrO_2$: $M_0 = 446\ G$; $K_a = 0.25\ Merg/cm^3$ [77]; $BaZnFe_{17}O_{27}$: $M_0 = 382\ G$; $K_a = 0.21\ Merg/cm^3$ [77]; $Fe_3O_4$: $M_0 = 477\ G$; $K_a = 0.135\ Merg/cm^3$ [76]; γ-$Fe_2O_3$: $M_0 = 374 G$; $K_a = 0.046\ Merg/cm^3$ [76]. $A_{max}$ and $h_{c,max}$ are maximum values of $A$ and $h_c$. The solid and dashed lines represent numerical (Eq. (31)) and analytical (Eq. (41)) calculations, respectively.

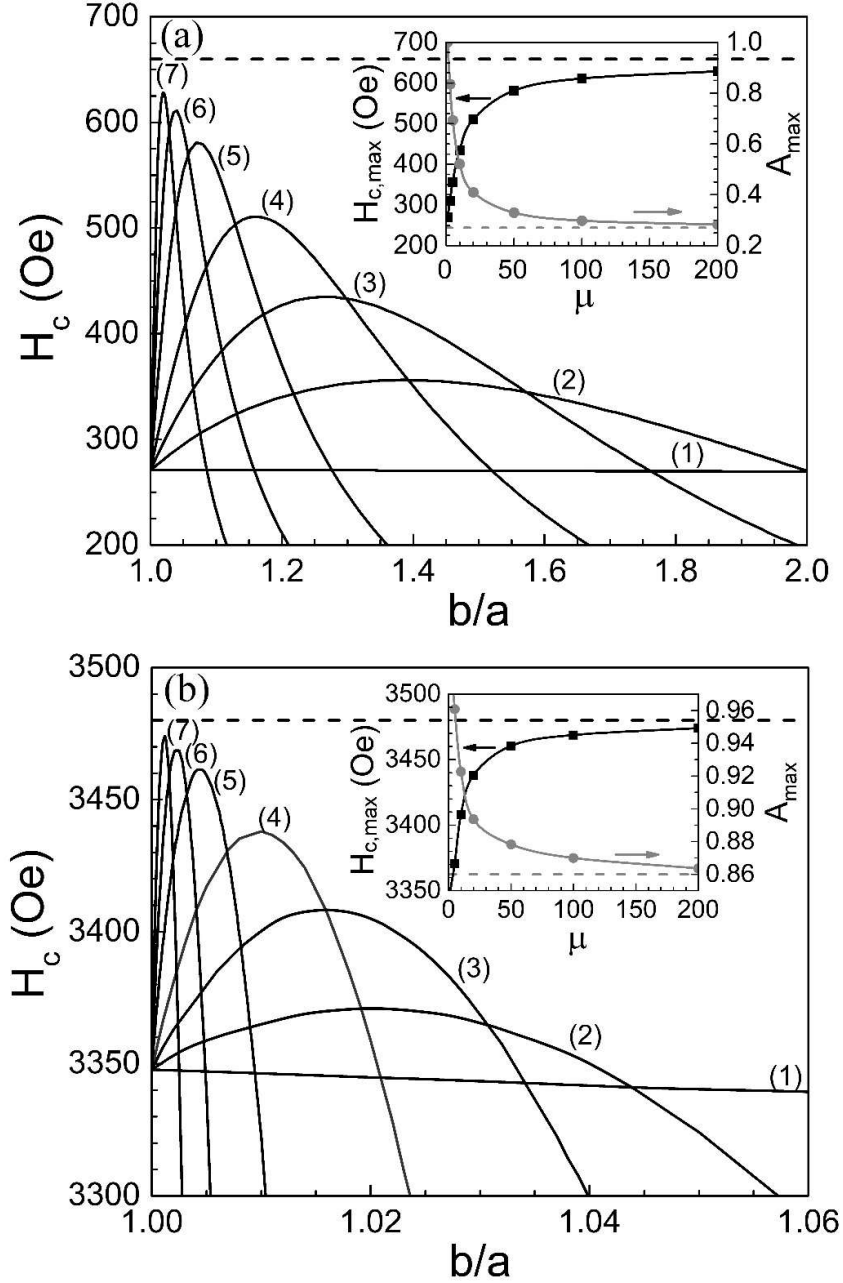

FIG. 6. The magnetic coercivity $H_c$ as a function of ratio $b/a$ for different values of the shell magnetic permeability μ in the case of (a) α-Fe core ($M_0 = 1.7$ κ$G$; $K_a = 0.48\ Merg/cm^3$[74]) and (b) Co core ($M_0 = 1.4\ kG$; $K_a = 5\ Merg/cm^3$[74]). The lines (1) – (7) were calculated by using $\mu$: (1) 1.1; (2) 5; (3) 10; (4) 20; (5) 50; (6) 100; (7) 200. The limiting value of the coercive force is (a) 660 Oe for α-Fe core, (b) 3480 Oe for Co core. The insets show dependencies $A_{max}$ (left axis) and $H_{c,max}$ (right axis) on $\mu$. The solid lines represent numerical (Eq. (31)) calculations.

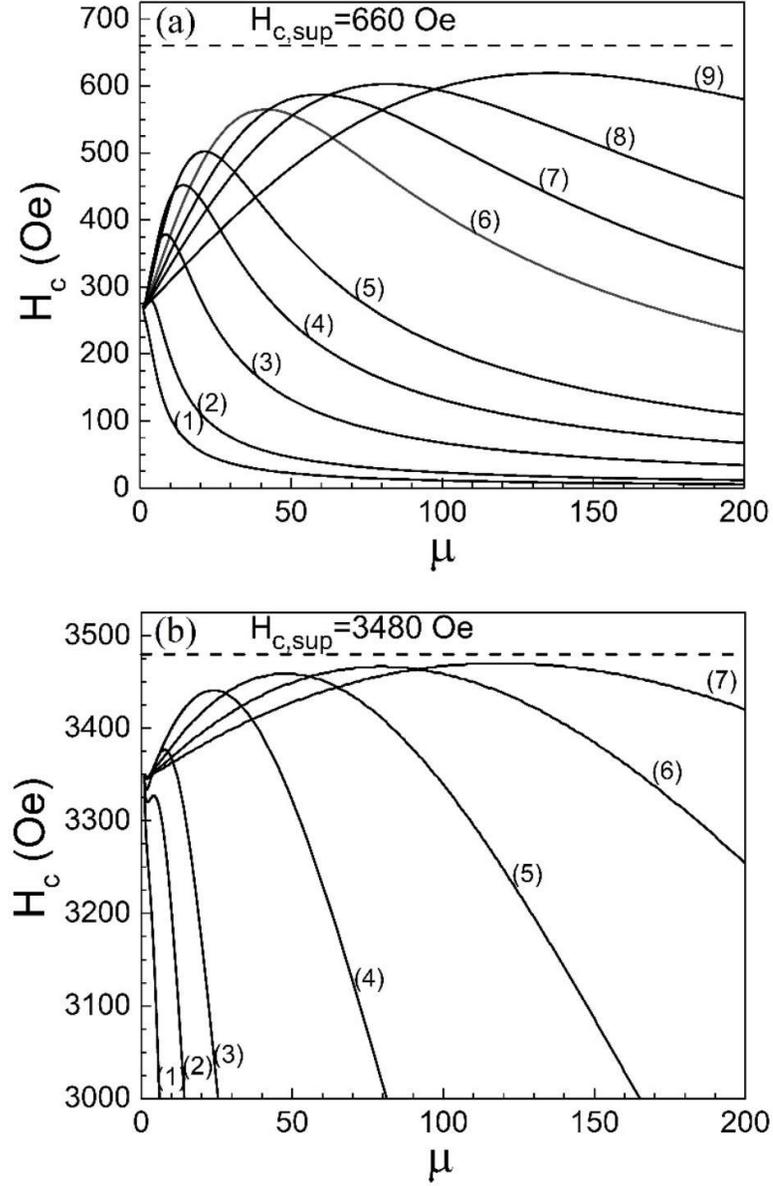

FIG. 7. The magnetic coercivity $H_c$ as a function of shell magnetic permeability μ for different values of ratio $b/a$ in the case of (a) α-Fe core ($M_0 = 1.7 kG$; $K_a = 0.48\ Merg/cm^3$[74]) and (b) Co core ($M_0 = 1.4 kG$; $K_a = 5\ Merg/cm^3$[74]). (a) lines (1) – (9) were calculated by using $b/a$: (1) 2.5; (2) 2.0; (3) 1.5; (4) 1.3; (5) 1.2; (6) 1.1; (7) 1.07; (8) 1.05; (9) 1.03. (b) lines (1) – (7) were calculated by using $b/a$: (1) 1.1; (2) 1.05; (3) 1.03; (4) 1.01; (5) 1.005; (6) 1.003; (7) 1.002. The limiting value of the coercive force is 660 Oe for α-Fe core and 3480 Oe for Co core. The solid lines represent numerical (Eq. (31)) calculations.

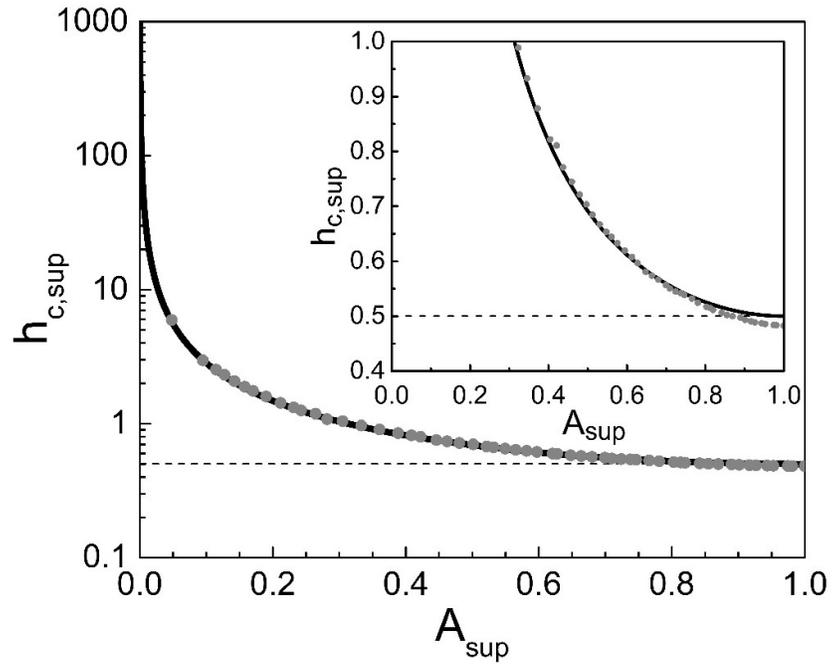

Fig.8 The dimensionless supremum coercive force $h_{c,sup}$ as a function of $A_{sup}$ (universal dependence). Dotted and solid lines represent numerical (Eq. (31), different values of $K_a$ and $M_0$) and analytical calculations (Eq. (46), (48)), respectively.

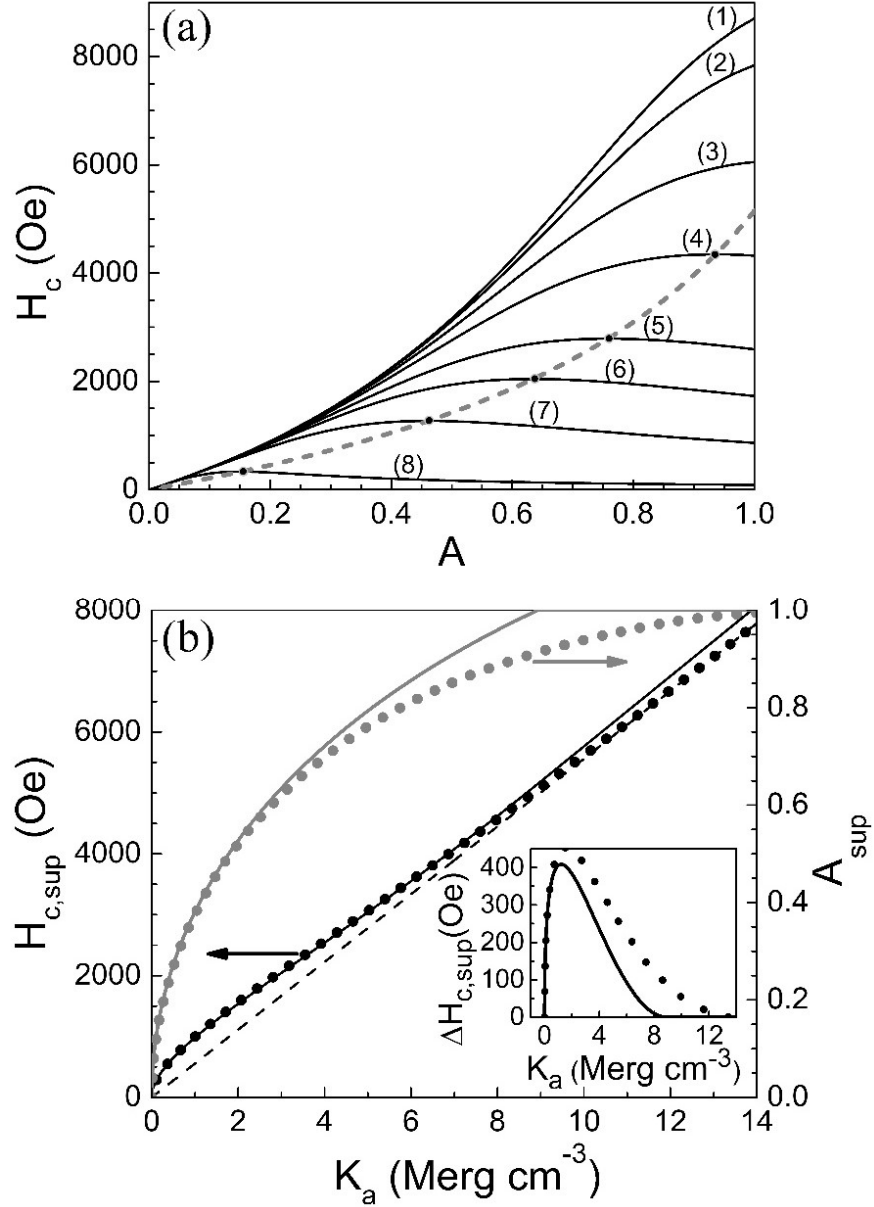

FIG. 9. (a) The limiting dependences of coercivity $H_c$ on $A$ for different $K_a$ and fixed $M_0 = 1.7\ kG$. The lines (1) – (9) were calculated by using $K_a\ (in\ Merg/cm^3)$: (1) 15.1; (2) 13.6; (3) 10.5; (4) 7.5; (5) 4.5; (6) 3.0; (7) 1.5; (8) 0.15. The dashed line corresponds to the dependence $H_{c,sup}(A_{sup})$ and intersects the limiting curves at the maximum points. (b) The dependence of the limiting coercivity $H_{c,sup}$ on $K_a$ (left axis) and the dependence of $A_{sup}$ on $K_a$ (right axis). The dashed line corresponds to $H_{c,sw}(K_a)$ dependence for an ordinary SW particle without a shell. The inset shows $\Delta H_{c,sup} = H_{c,sup} - H_{c,sw}$ as a function of $K_a$. The solid lines and dots represent analytical (Eqs. (41), (43), (44), (46), (48)) and numerical (Eq. (31)) calculations, respectively.

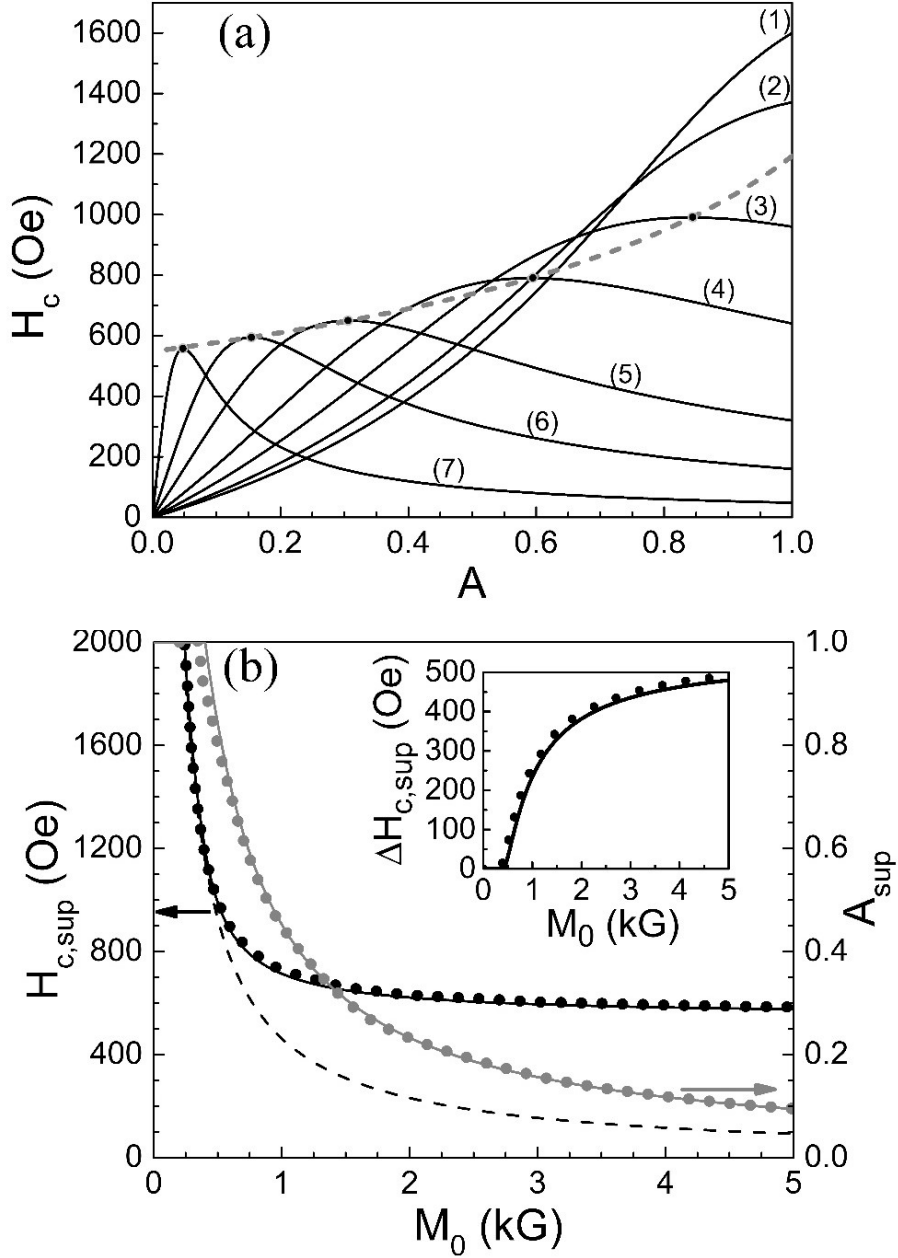

FIG. 10. (a) The limiting dependences of coercivity $H_c$ on A for different $M_0$ and fixed $K_a = 0.48\ Merg/cm^3$. The lines (1) – (9) were calculated by using $M_0$ (in G): (1) 300; (2) 350; (3) 500; (4) 750; (5) 1500; (6) 3000; (7) 10000. The dashed line corresponds to the dependence $H_{c,sup}(A_{sup})$ and intersects the limiting curves at the maximum points. (b) The dependence of the limiting coercivity $H_{c,sup}$ on $M_0$ (left axis) and the dependence of $A_{sup}$ on $M_0$ (right axis). The dashed line corresponds to $H_{c,sw}(M_0)$ dependence for an ordinary SW particle without a shell. The inset shows $\Delta H_{c,sup} = H_{c,sup} - H_{c,sw}$ as a function of $M_0$. The solid lines and dots represent analytical (Eqs. (41), (43), (44), (46), (48)) and numerical (Eq. (31)) calculations, respectively.

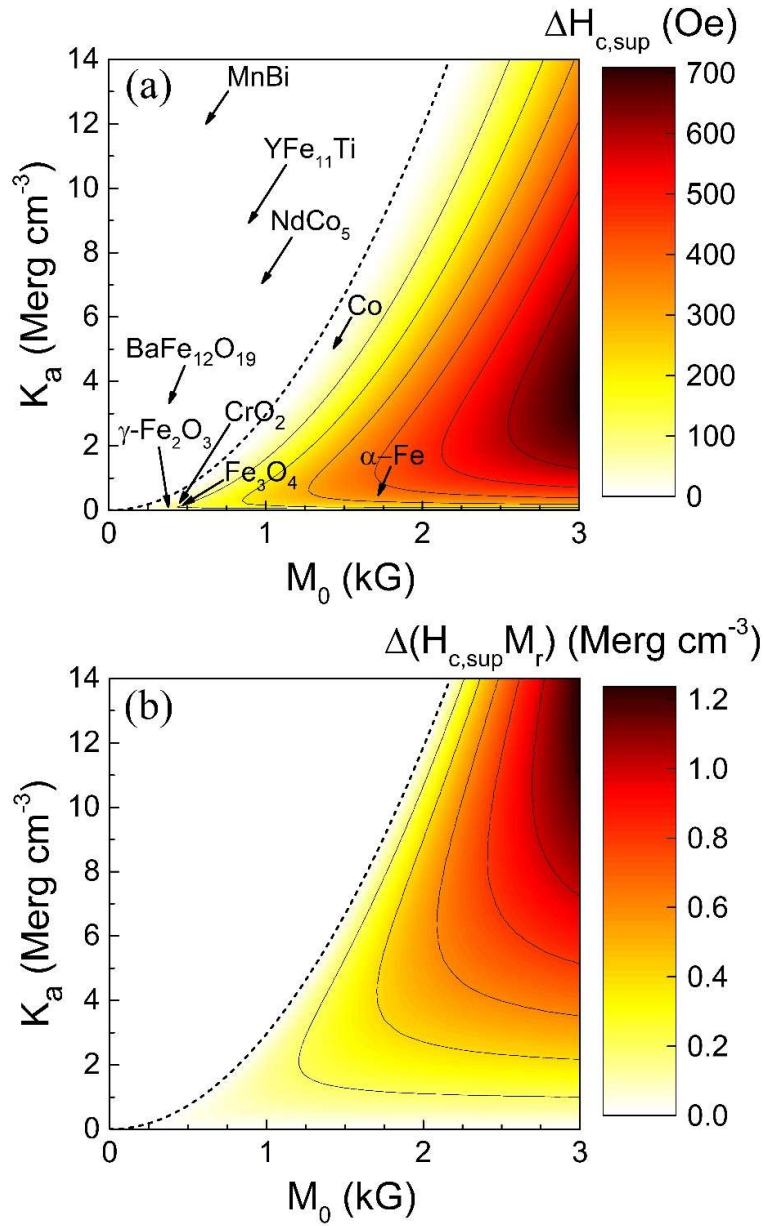

Fig.11. (a) $\Delta H_{c,sup} = H_{c,sup} - H_{c,sw}$ versus $M_0$ and $K_a$. The following materials are marked with arrows ($M_0$ in $kG$; $K_a$ in $Merg/cm^3$): α-Fe (1.71; 0.48) [74]; Co (1.44; 5.0) [74]; $CrO_2$ (0.446; 0.25) [77]; $Fe_3O_4$ (0.477; 0.135) [76]; γ-$Fe_2O_3$ (0.374; 0.046) [76]; $BaFe_{12}O_{19}$ (0.382; 3.3) [77]; $NdCo_5$ (0.979; 7.0) [77]; $YFe_{11}Ti$ (0.891; 8.9) [77]; MnBi (0.621; 12.0) [77]. (b) $\Delta(H_{c,sup}M_r) = H_{c,sw}M_{r,sw} - H_{c,sup}M_r$ versus $M_0$ and $K_a$. The dotted line delimits the area of the coercive force enhancement (Eq. (49)). The graphs were plotted by using Eq. (41).